\documentclass[9pt,twocolumn,twoside]{osajnl}
\usepackage{upgreek}

\journal{ol} 

\setboolean{shortarticle}{true}

\title{Fine Features of Optical Potential Well Induced by Nonlinearity}

\author[1,*]{Lei-Ming Zhou}
\author[2,3]{Yaqiang Qin}
\author[4]{Yuanjie Yang}
\author[2,3]{Yuqiang Jiang}
\affil[1]{Department of Electrical and Computer Engineering, National University
of Singapore, Singapore 117583, Singapore}
\affil[2]{State Key Laboratory of Molecular Developmental Biology, Institute of Genetics and Developmental Biology, Chinese Academy of Sciences, Beijing, 100101, China}
\affil[3]{University of Chinese Academy of Sciences, Beijing, 100049, China}
\affil[4]{School of Physics, University of Electronic Science and Technology
of China, Chengdu 610054, China}
\affil[*]{Corresponding author: leiming.zhou@nus.edu.sg}

\dates{Compiled November 12, 2020}

\ociscodes{(350.4855) Optical tweezers or optical manipulation; (190.3970) Microparticle nonlinear optics; (320.2250) Femtosecond phenomena.}

\doi{\url{http://dx.doi.org/10.1364/XX.XX.XXXXXX}}

\begin{abstract}
Particles trapped by optical tweezers, behaving as mechanical oscillators in optomechanical system, have found tremendous applications in various disciplines and are
still arousing research interests in frontier and fundamental physics.
These optically trapped oscillators provide compact particle confinement
and strong oscillator stiffness. But these features are limited by
the size of the focused light spot of laser beams, which is typically
restricted by the optical diffraction limit. Here, we propose to build optical potential well with fine features assisted by the nonlinearity of the particle material, which is independent of the
optical diffraction limit. We show that the potential well shape can
have super-oscillation-like features and Fano-resonance-like phenomenon,
and the width of optical trap is far below the diffraction limit.
The particle with nonlinearity trapped by ordinary optical beam provides
a new platform with sub-diffraction potential well and can have applications in high accuracy optical manipulation and high precision metrology.
\end{abstract}

\setboolean{displaycopyright}{true}

\begin{document}

\maketitle

\emph{\label{sec:Introduction}}Optical tweezers have found tremendous
of applications in various areas, including colloidal sciences, chemistry
and biophysics since its invention in several decades ago \cite{AshkinOL1986,GrierNature2003,DholakiaRMP2010Colloquium,QiuLSA2018More}.
Especially, trapping with focused laser beams has advantages of non-contact,
non-clamping and tunability. They are still arousing research interests
in frontier and fundamental physics \cite{LiTCScience2010,LiTCNPhysics2011Millikelvin,YinIJMPB2013optomechanics,ZhouLPR2017optical,GeraciPRL2010Short,ArvanitakiPRL2013Detecting,MartinezScience2016,GieselerNNano2014Dynamic},
such as obtaining negative pulling force \cite{DingAP2019PhotonicTractor,LiAOP2020opticalpulling},
searching non-Newtonian force \cite{GeraciPRL2010Short}, detecting
gravitational wave \cite{ArvanitakiPRL2013Detecting}, building quantum
Carnot engine \cite{MartinezScience2016} and investigating the thermodynamics
of non-equilibrium steady system \cite{GieselerNNano2014Dynamic}.
In these researches, manipulating the particles in micro/nano-scale
plays a key role. Trapping and manipulating particles in micro/nano-scale
provide compact confinement, thus strong stiffness and high frequency
of the optical trapped oscillator. These features make high sensitivity
metrology and the observation of micro-scale dynamics possible \cite{LiTCScience2010,ZhouOL2018Displacement}.

However, the size of the focused light spot of laser beams is typically
restricted by the optical diffraction limit \cite{JonesBook2015optical}. Light spot (or the trapping potential well) with sub-diffraction size means that we can trap two particles away from each other while keeping the distance between them below the diffraction limit. This provides a way for parallel manipulation with compact size (particle arrangement below sub-diffraction at the same time) and high precision manipulation. Also, the width of the potential well will affect the stiffness of the trapped oscillator in optomechanical systems. Higher stiffness will increase the oscillation frequency and the sensitivity of the metrology applications based on the trapped oscillator.
In order to obtain the sub-diffraction potential well, super-oscillation
beams has been proposed to get smaller light spot \cite{SinghLSA2017optical,ChenLSA2019SuperOscillation,NagarOL2019}. Though super-oscillation beam can have a spot much smaller than the wavelength, the spot takes only a very minor part of the total incident power, which reduces the efficiency.

In those research above, the trapped particle or the oscillator is
made of optical linear materials. Here, we investigate the trapping behaviors of particles with various high-order nonlinearities and report the building of a narrow potential well with width far below the diffraction limit. The proposal is based on
the trapping theory of nonlinear particles. And recently, the effect induced by
particle nonlinearity has been noticed with the introduction of the
femtosecond light pulse \cite{JiangNPhysics2010,ZhangNL2018Nonlinearity-Induced,HuangNanophotonics2020}.
Though the nonlinear coefficient of the particle is small, the
strong light intensity of femtosecond laser beam makes the
nonlinearity induced effect observable. The theory here then can be verified with the developing experiment techniques in the near future.

This paper is organized as follows. First, we describe the system
investigated and the method used. Second, we show that different orders of nonlinearity exert different effects
on the system. The potential well function can have Fano-resonance-like phenomenon and super-oscillation-like fine features. Then, we discuss
the the experiment realization and validity of our theory.

\begin{figure}
\begin{center}
\includegraphics[width=7cm]{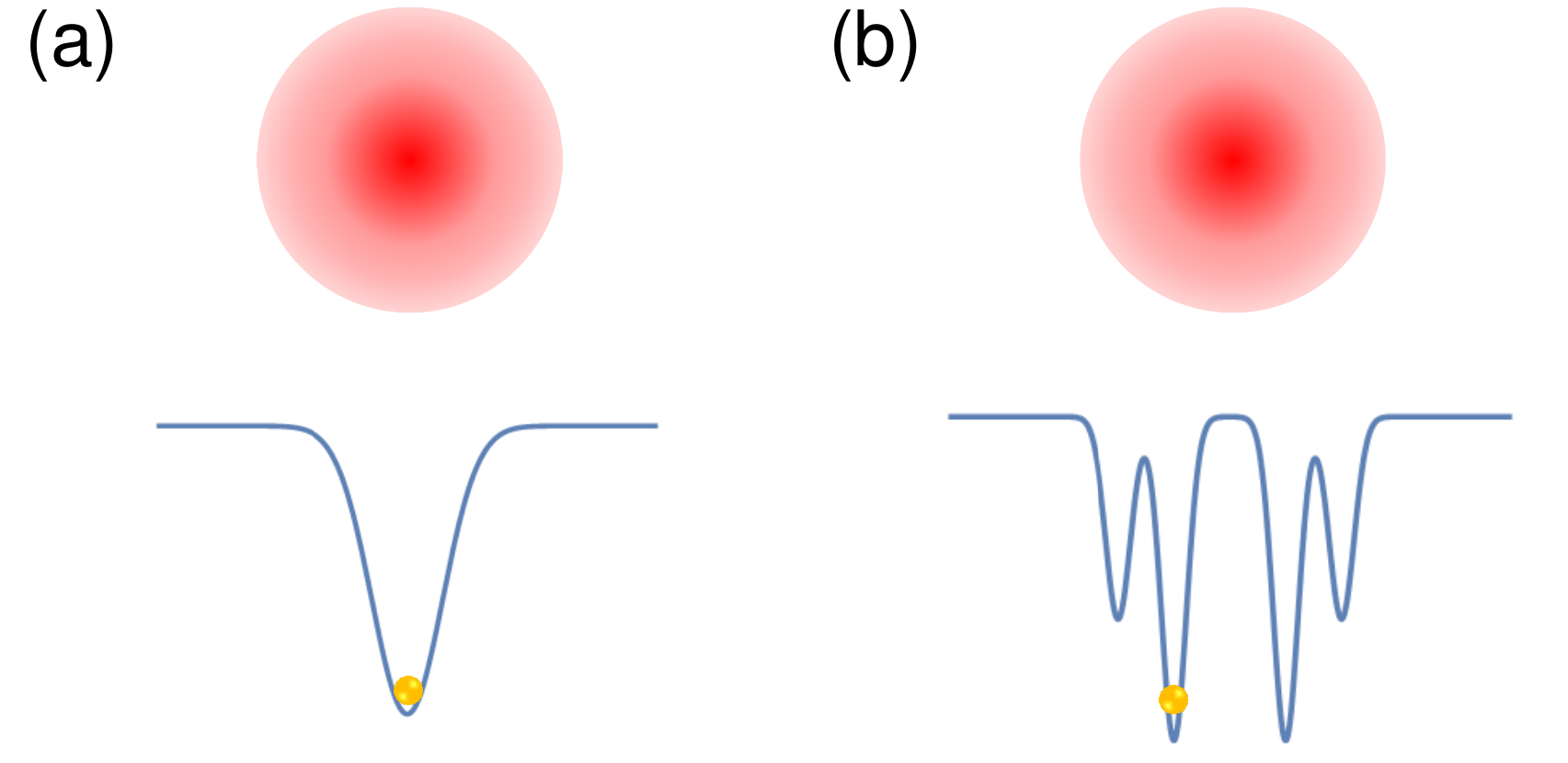}
\caption{Illustration of the optical potential well for (a) linear
particle and (b) nonlinear particle in a single light spot.\label{fig:illustration-of-system}}
\end{center}
\end{figure}

\textcolor{blue}{\label{sec:Setup}}Typically, gradient force generated by a light spot is used in optical trapping of micro/nano
particles. Without loss of generality, we use a linear polarized Gaussian
laser beam to represent this light spot, where the electromagnetic field has
transverse distribution of the form $E(\rho)=|\boldsymbol{{\rm E}}|=E_{0}e^{-\rho^{2}}$ . The radial distance from the center is denoted by $\rho$ and the
amplitude of the electric field in the center is $E_0$ . The wave-vector of the incident beam is $k_{\rm{m}}$ in the particle's surrounding medium, where $k_{\rm{m}}=n_{\rm{m}}k_{0}=2\pi n_{\rm{m}}/\lambda_0$ 
and the wavelength in vacuum is $\lambda_0$.  The refractive index of the particle and the surrounding medium are $n_{\rm{p}}$ and $n_{\rm{m}}$ respectively. For a small spherical particle with radius $a$  fulfilling the Rayleigh approximation (i.e., $k_{\rm{m}} a\ll{1}$ and $n_{\rm{p}}{k_0} a\ll {1}$), the gradient force exerted by the electromagnetic field is $F=\frac{1}{4}\varepsilon_{\rm{m}}\varepsilon_{0}{\rm Re}(\alpha)\nabla|\boldsymbol{{\rm E}}|^{2}$. Here, $\alpha=\alpha_0/(1-i {{k^3_{\rm m}}}{\alpha_0}/6\pi\varepsilon_{\rm{m}}) $ is the effective polarizability of the particle, $\alpha_0=4\pi a^{3}\frac{\varepsilon_{\rm{r}}-1}{\varepsilon_{\rm{r}}+2}$ is the static polarizability \cite{JonesBook2015optical,SvobodaOL94Optical}
and ${\varepsilon_{\rm{r}}}=\varepsilon_{\rm{p}}/\varepsilon_{\rm{m}}$ is the relative refractive index. The
phenomenological potential function is defined as $U(\rho)=-\int F(\rho){\rm d}{\rho}$. It's noted here that we consider the gradient force only to make the analysis of the trapping mechanism clearer. In the experiment, the scattering force should be eliminated by a counter-propagating beam. Otherwise, the analysis here is valid only on the radial direction on the transverse plane of the beam waist. For a particle with linear optical material, the polarizability is
independent of the location, thus we have $U=-\frac{1}{4}\varepsilon_{m}\varepsilon_{0}{\rm Re}(\alpha)|\boldsymbol{{\rm E}}|^{2}$. The shape of the trapping potential thus is the same
with the profile of the light spot, which has a size limited by the
diffraction limit (illustrated qualitatively by Fig. \ref{fig:illustration-of-system}(a)).

However, for a particle with nonlinear optical material, the effective
polarizability $\alpha$ depends on $E$ (thus also dependent on particle
location $\rho)$. The susceptibility of the particle material with
high order nonlinearity in a light spot is location dependent and
reads
\begin{equation}
\chi(\rho)=\chi^{(1)}+\chi^{(3)}E^{2}+\chi^{(5)}E^{4}+....\label{eq:highOrderChi}
\end{equation}
Considering a field distribution with Gaussian shape in its transverse
plane as before, the permittivity ($\varepsilon=1+\chi$) of the particle
is also location dependent and reads
\begin{equation}
\varepsilon(\rho)=\varepsilon_{1}+\Delta\varepsilon_{3}e^{-2\rho^{2}}+\Delta\varepsilon_{5}e^{-4\rho^{2}}+....\label{eq:highOrderepsilon}
\end{equation}
Among Eq. (\ref{eq:highOrderepsilon}), $\varepsilon_1=1+\chi^{(1)}$ and different orders of nonlinear coefficient cause different permittivity changes and are denoted as
$\Delta\varepsilon_{i}=\chi^{(i)}E_{0}^{i-1}$ ($i=3,5,7,\cdots$)
respectively. And finally, the effective polarizability $\alpha$
of the particle is location dependent, which induces changes in the
potential function as illustrated qualitatively in Fig. \ref{fig:illustration-of-system}(b).

Though the accurate potential function can be calculated with
numerical integration of force $F$, we
provide an approximate analytical expression to investigate it first.
Suppose the location dependence of the permittivity $\varepsilon(\rho)$
is much less than that of the electrical field and the polarizability $\alpha\approx \alpha_0$ for a Rayleigh particle, the approximate
potential function thus is
\begin{equation}
U_{\rm approx}=-c_{1}{\rm Re}(\frac{\varepsilon_{\rm r}(\rho)-1}{\varepsilon_{\rm r}(\rho)+2})|\boldsymbol{{\rm E}}|^{2}.\label{eq:UApprox}
\end{equation}
Among Eq. (3), $c_{1}=\pi a^{3}\varepsilon_{\rm{m}}\varepsilon_{0}$ is independent
of the particle location and we takes it as a normalization coefficient of the potential function. This approximate expression is different from the accurate numerical result sometimes as shown later, because we have supposed the location dependence of the permittivity $\varepsilon(\rho)$
is much less than that of the electrical field. But it provides us a guidance to predict the main behaviors of the potential-well. Now, we show the results for various order of nonlinearity and their
effects. Without loss of generality, it is supposed that the permittivity
of the surrounding medium $\varepsilon_{\rm{m}}$ equals 1 (thus $\varepsilon_{\rm{r}}=\varepsilon_{\rm{p}}$)
below.
\begin{figure}
\begin{center}
\includegraphics[width=8.5cm]{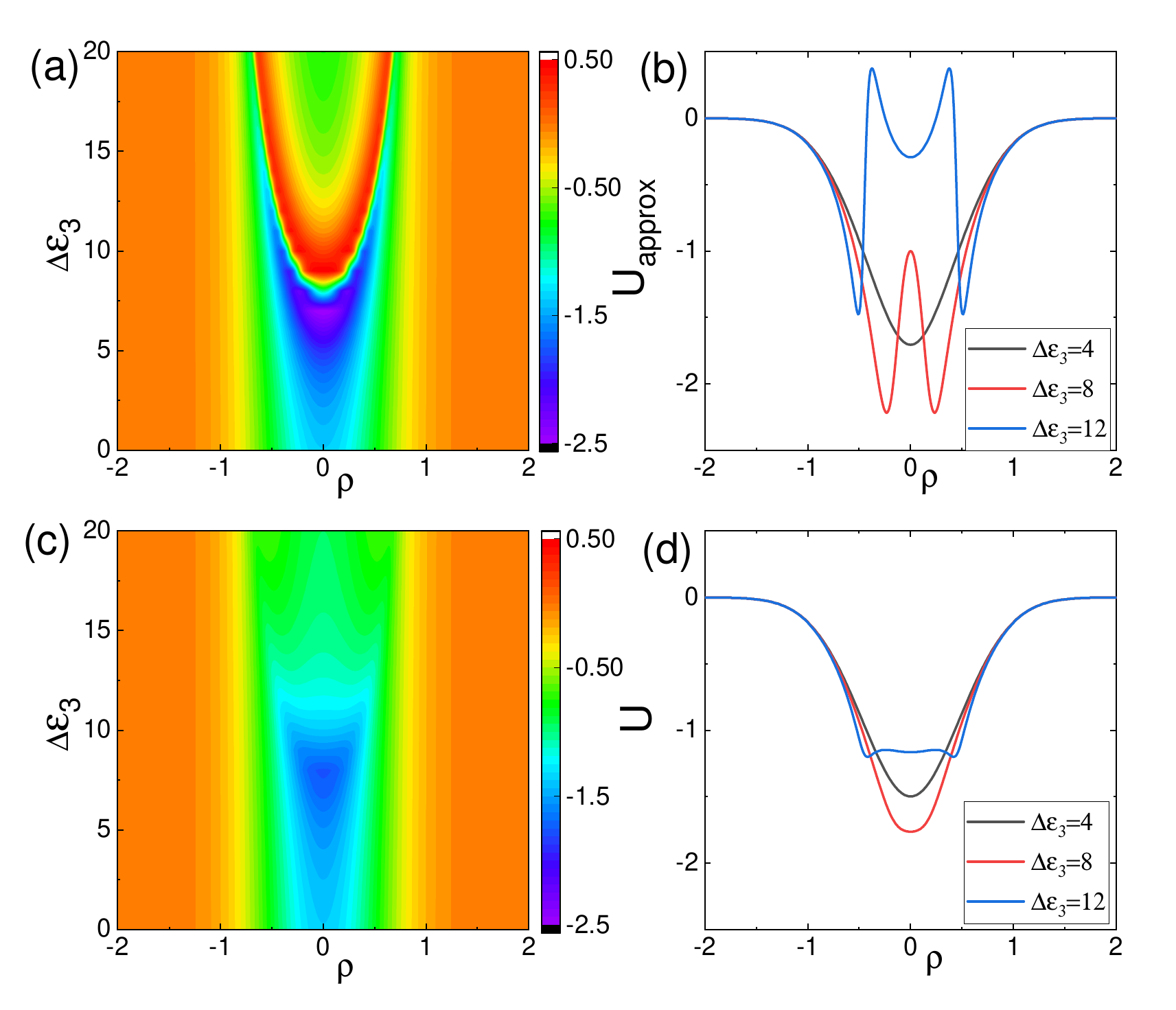}
\caption{The normalized potential for various nonlinearity when only the
third order nonlinearity $\chi^{(3)}$ is included. (a) Approximate explicit potential $U_{\rm approx}$ for
different $\Delta\varepsilon_{3}$ when $\varepsilon_{1}^{'}=-10$
and $\varepsilon_{1}^{''}=1.0$. (b) Potential well
shape for $U_{\rm approx}$ with several different $\Delta\varepsilon_{3}$. (c-d) The same
as Figs. (a-b) for accurate potential $U$  calculated with numerical integration.\label{fig:Ur-chi3}}
\end{center}
\end{figure}
\begin{figure}
\begin{center}
\includegraphics[width=8.5cm]{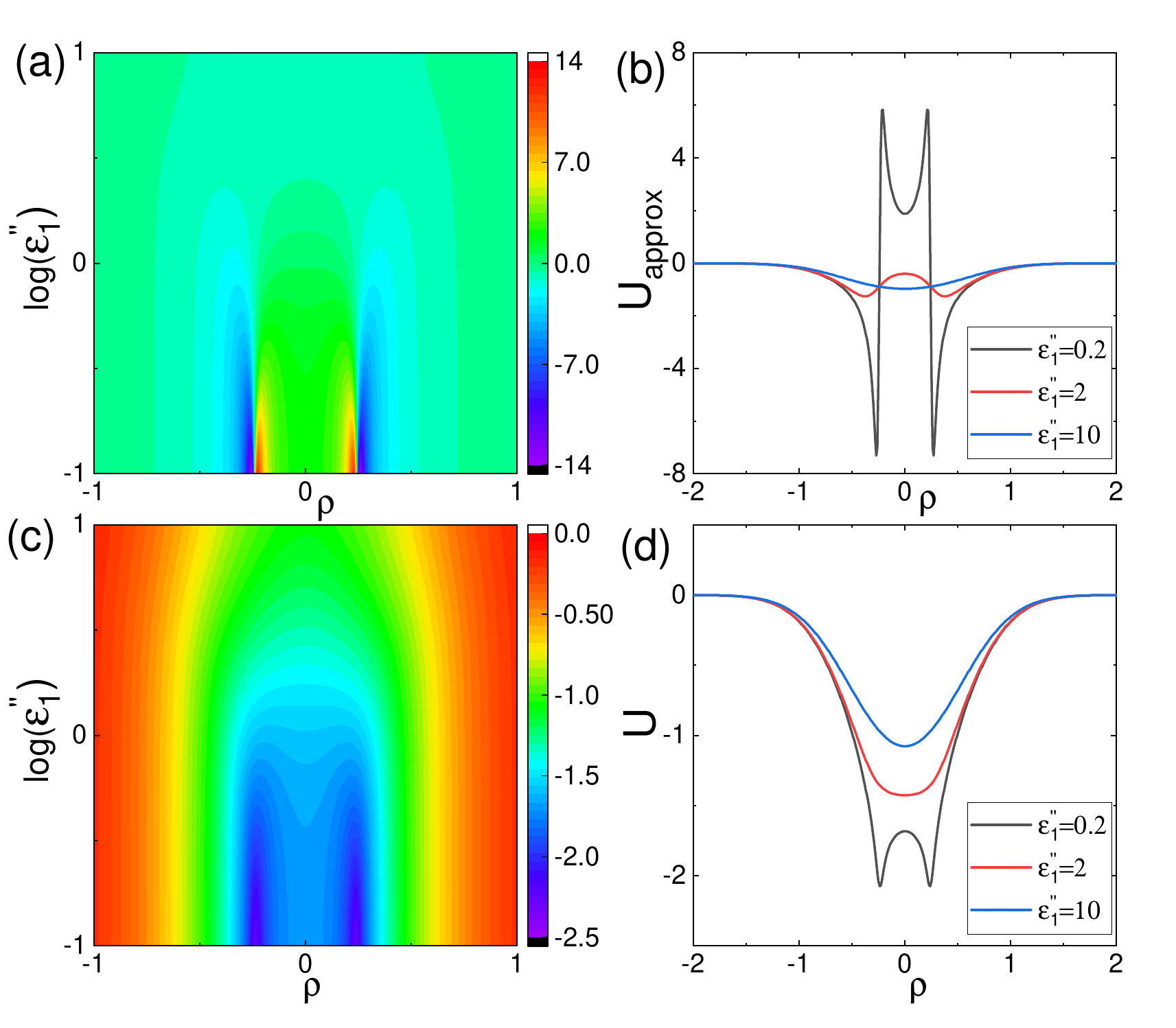}
\caption{The normalized potential for various absorption when only the
third order nonlinearity $\chi^{(3)}$ is included. (a) Approximate explicit potential $U_{\rm approx}$ for
different $\varepsilon_{1}^{''}$ when $\varepsilon_{1}^{'}=-10$
and $\Delta\varepsilon_{3}=9$. (b) Potential well
shape for $U_{\rm approx}$ with several different $\Delta\varepsilon_{1}^{''}$. (c-d) The
same as (a-b) for accurate potential $U$  calculated with numerical integration.\label{fig:Ur-e1pp}}
\end{center}
\end{figure}

For particles with only the third order
nonlinearity coefficient $\chi^{(3)}$, the permittivity change is
$\Delta\varepsilon=\Delta\varepsilon_{3}e^{-2\rho^{2}}=\chi^{(3)}E_{0}^{2}e^{-2\rho^{2}}$.
The particle's linear relative permittivity is $\varepsilon_{1}=\varepsilon_{1}^{'}+i\varepsilon_{1}^{''}$,
where the real and imaginary part of the permittivity are denoted with
$\varepsilon_{1}^{'}$ and $\varepsilon_{1}^{''}$ respectively. The
normalized potential function thus reads
\begin{align}
U_{\rm approx} & =-{\rm Re}(\frac{\varepsilon_{1}^{'}+i\varepsilon_{1}^{''}+\Delta\varepsilon_{3}e^{-2\rho^{2}}-1}{\varepsilon_{1}^{'}+i\varepsilon_{1}^{''}+\Delta\varepsilon_{3}e^{-2\rho^{2}}+2})e^{-2\rho^{2}}\nonumber \\
 & =-e^{-2\rho^{2}}+\frac{3e^{-2\rho^{2}}(\varepsilon_{1}^{'}+\Delta\varepsilon_{3}e^{-2\rho^{2}}+2)}{(\varepsilon_{1}^{'}+\Delta\varepsilon_{3}e^{-2\rho^{2}}+2)^{2}+\varepsilon_{1}^{''2}}\nonumber \\
 & \triangleq U_{{\rm linear}}+U_{{\rm nonlinear}}^{(3)}.
\end{align}

\begin{figure}
\begin{center}
\includegraphics[width=8cm]{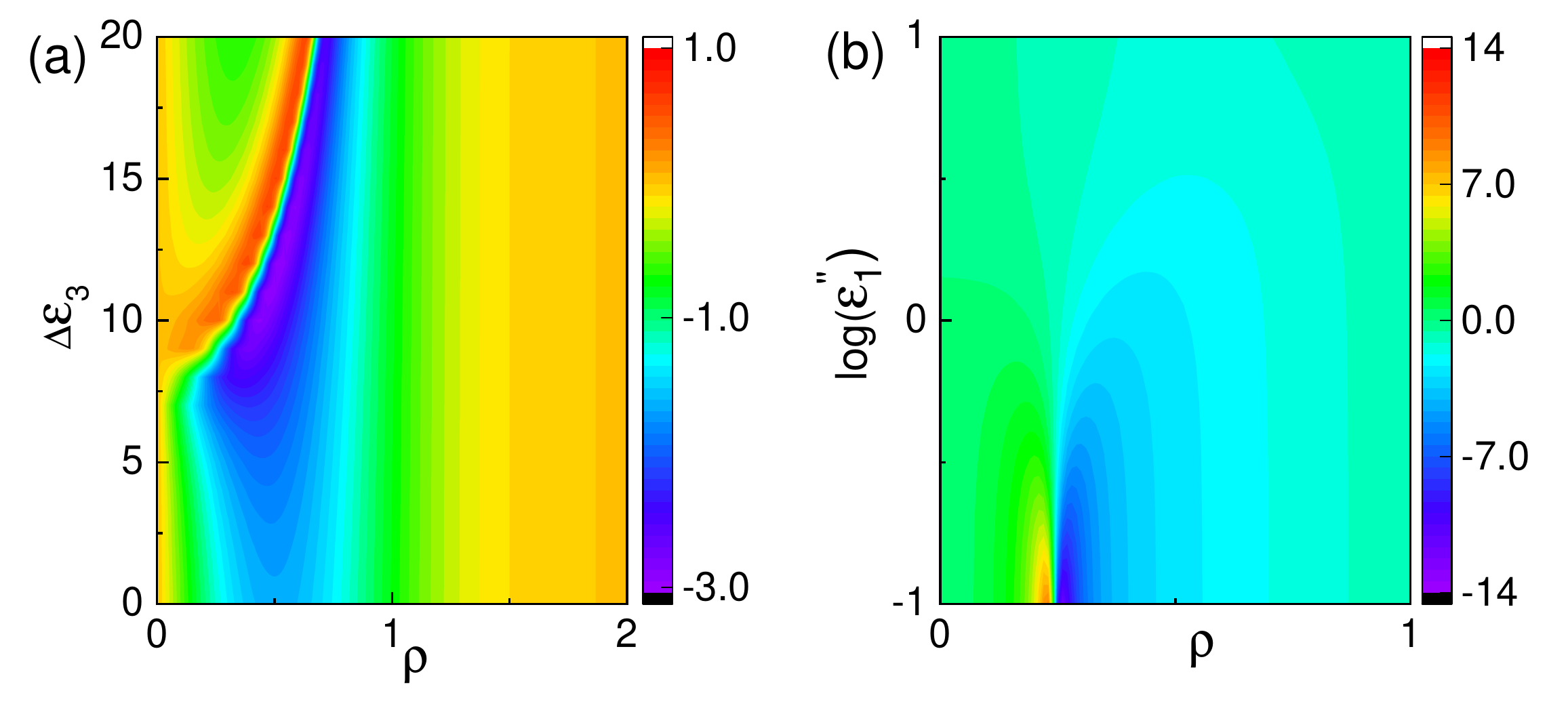}
\caption{(a) The normalized force $F(\rho)$ for different $\Delta\varepsilon_{3}$
when $\varepsilon_{1}^{'}=-10$ and $\varepsilon_{1}^{''}=1.0$. (b) The normalized force $F(\rho)$ for different $\varepsilon_{1}^{''}$
when $\varepsilon_{1}^{'}=-10$ and $\Delta\varepsilon_{3}=9$. \label{fig:Fr-chi3}}
\end{center}
\end{figure}

It can be seen that the linear potential function $U_{{\rm linear}}$
keeps the Gaussian shape, while the nonlinear potential $U_{{\rm nonlinear}}^{(3)}=3e^{-2\rho^{2}}U_{{\rm core}}^{(3)}$
shows a Fano-like shape core with a Gaussian envelop, as shown in
Fig. \ref{fig:Ur-chi3}(a-b). When $\Delta\varepsilon_{3}$ (proportional
to $\chi^{(3)}$) increases, the potential function changes from single
potential well to double potential well, and even get a third potential
well in the center (when $\Delta\varepsilon_{3}=12$ as shown in Fig.
\ref{fig:Ur-chi3}(a)). Accurate potential well $U$ got from the
numerical integration of force $F$ arrives the similar result as
shown in Fig. \ref{fig:Ur-chi3}(c-d). The potential well changes from single well to double well as predicted, and the location of the double well coincides with the approximation result for cases where the $\Delta\varepsilon_{3}$ is away from the just splitting point. The core function $U_{{\rm core}}^{(3)}=\frac{\varepsilon_{1}^{'}+\Delta\varepsilon_{3}e^{-2\rho^{2}}+2}{(\varepsilon_{1}^{'}+\Delta\varepsilon_{3}e^{-2\rho^{2}}+2)^{2}+\varepsilon_{1}^{''2}}$
shows its maximum/minimum values $\pm\frac{1}{2\varepsilon_{1}^{''}}$
when $\varepsilon_{1}^{'}+\Delta\varepsilon_{3}e^{-2\rho^{2}}+2=\pm\varepsilon_{1}^{''}$. These points are also the points of the maximum/minimum of the Fano-like shape of potential function. The depth and width of the potential well is different from the approximate analytical result, which is mainly affected by the imaginary part of the permittivity  $\varepsilon_{1}^{''}$ as we will show below. We also show the corresponding normalized force $F$ for various $\Delta\varepsilon_{3}$ in Fig. \ref{fig:Fr-chi3}(a) as a reference. In the numerical calculation, $a=30\ {\rm nm}$ and $\lambda_0=840\ {\rm nm}$. However, as long as the Rayleigh approximation is fulfilled, the radius of the particle doesn't affect the shape of the normalized potential well in fact.

While $\Delta\varepsilon_{3}$ determines the location of the transition
point, the absorption part $\varepsilon_{1}^{''}$ mainly affects the width
and depth of the potential well. We show these behaviors of the shape of the potential well in Fig. \ref{fig:Ur-e1pp}(a-b) for various $\varepsilon_{1}^{''}$.
It can be seen that when the absorption $\varepsilon_{1}^{''}$  is small, the potential well depth increases very much and the Fano-like shape is pronounced. Large  $\varepsilon_{1}^{''}$ (i.e., large particle absorption) will flat the potential curve and even change the double well into single well. Numerical result shows similar behaviors and the flatting effect of the imaginary part of the permittivity is more obvious. The normalized force $F$ for various  $\varepsilon_{1}^{''}$ is also shown in Fig. \ref{fig:Fr-chi3}(b).

When the permittivity is real, the simplified potential reads
\begin{equation}
U_{0}=-(1-\frac{3}{\varepsilon_{1}+2+\Delta\varepsilon_{3}e^{-2\rho^{2}}})e^{-2\rho^{2}}.\label{eq:U0shapeApprox}
\end{equation}
Among Eq. (\ref{eq:U0shapeApprox}), there is a singular point when $\varepsilon_{1}+2+\Delta\varepsilon_{3}e^{-2\rho^{2}}=0$. It is corresponding to the maximum/minimum point of the double-well potential function as mentioned above when  $\varepsilon_{1}^{''}$ is very small. This point has been experimentally observed before \cite{JiangNPhysics2010}, where they used the gold nano-particle. For metallic particle
such as gold particle, $\varepsilon_{1}+2<0$ and $\Delta\varepsilon_{3}e^{-2\rho^{2}}>0$.

\begin{figure}
\begin{center}
\includegraphics[width=8cm]{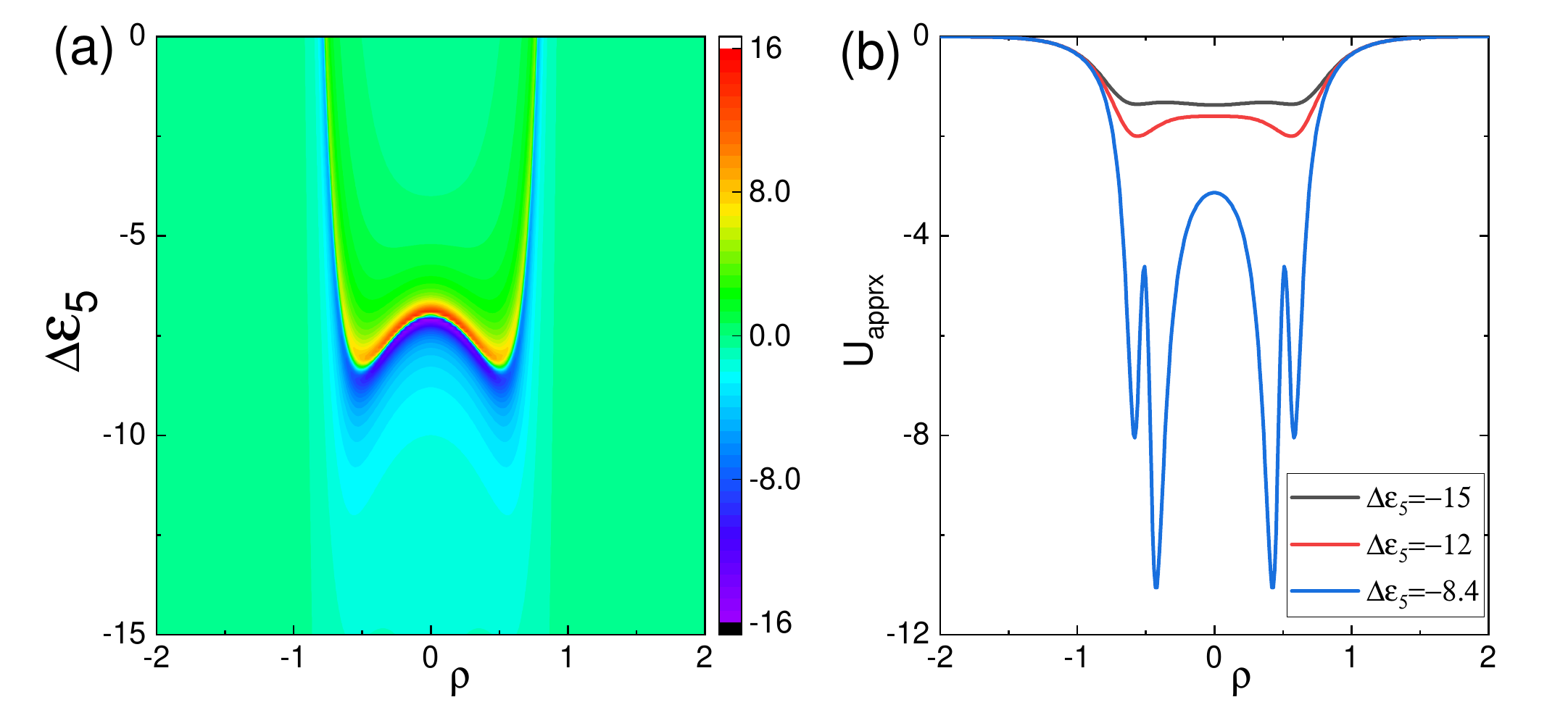}
\caption{(a) The normalized potential function $U_{\rm approx}$ for various $\Delta\varepsilon_{5}$
when $\varepsilon_{1}=-5+0.1i$ and $\Delta\varepsilon_{3}=10$.
(b) Several representative potential-well function for different $\Delta\varepsilon_{5}$. Higher-order nonlinearity can induce obvious super-oscillation-like potential function such as  $\Delta \varepsilon_{5}=-8.4$.\label{fig:U0_chi5}}
\end{center}
\end{figure}

If higher order nonlinearity is included, they change the part $\Delta\varepsilon$
and the potential shape could have more details. Taking $\chi^{(5)}$
as an example, the normalized potential function reads

\begin{align}
U_{\rm approx} & =-{\rm Re}(\frac{\varepsilon_{1}^{'}+i\varepsilon_{1}^{''}+\Delta\varepsilon_{3}e^{-2\rho^{2}}+\Delta\varepsilon_{5}e^{-4\rho^{2}}-1}{\varepsilon_{1}^{'}+i\varepsilon_{1}^{''}+\Delta\varepsilon_{3}e^{-2\rho^{2}}++\Delta\varepsilon_{5}e^{-4\rho^{2}}+2})e^{-2\rho^{2}}\nonumber \\
 & =-e^{-2\rho^{2}}+\frac{3e^{-2\rho^{2}}(\varepsilon_{1}^{'}+\Delta\varepsilon_{3}e^{-2\rho^{2}}+\Delta\varepsilon_{5}e^{-4\rho^{2}}+2)}{(\varepsilon_{1}^{'}+\Delta\varepsilon_{3}e^{-2\rho^{2}}+\Delta\varepsilon_{5}e^{-4\rho^{2}}+2)^{2}+\varepsilon_{1}^{''2}}\nonumber \\
 & \triangleq U_{{\rm linear}}+U_{{\rm nonlinear}}^{(5)}.\label{eq:FanoShape-1}
\end{align}
The nonlinear potential $U_{{\rm nonlinear}}^{(5)}=3e^{-2\rho^{2}}U_{{\rm core}}^{(5)}$
also has a Gaussian envelop. If the sign of $\chi^{(5)}$ is the same
as that of $\chi^{(3)}$, it doesn't change the shape of the total
nonlinearity-induced permittivity $\Delta\varepsilon$ because $\Delta\varepsilon_{3}e^{-2\rho^{2}}$
and $\Delta\varepsilon_{5}e^{-4\rho^{2}}$ have the same monotonicity.
Thus the potential shape keeps the same as the case of $\chi^{(3)}$.
However, $\chi^{(5)}$with opposite sign will show more oscillations
as shown in Fig. \ref{fig:U0_chi5}(a-b). As shown in Fig. \ref{fig:U0_chi5}(b),
the potential function shows a super-oscillation-like behavior with chosen parameters. For
a strongly focused light spot, the total width of the Gaussian envelop
is limited by the diffraction limit and thus the single wells in this
super-oscillation-like potential well will have width far below the diffraction
limit.

More higher order nonlinearities cannot introduce more oscillations.
All the changed permittivity terms ($\Delta\varepsilon_{i}e^{-(i-1)\rho^{2}}$)
can be gathered together into two groups, according to the sign of
the nonlinearity coefficient. All the terms associated with positive/negative
high-order nonlinearities shows the same monotonicity when the location
of the particle changes. Then the behaviors induced by all the nonlinearities
are the same as the case where only $\chi^{(3)}$ and $\chi^{(5)}$
are included.

The inclusion of even-order nonlinearities shows the same behaviors
as we have found before. In the above discussion, we only consider
odd-order nonlinearities for simplification because the material of
the trapped particle is usually with central inversion symmetry. Materials
with this symmetry have only zero even-order nonlinearities. However,
even if we introduce even-order nonlinearities, they can be analyzed
as what we did for higher order nonlinearities. Through grouping positive
and negative nonlinearity coefficients, the total potential function
show the same behaviors as when only $\chi^{(3)}$ and $\chi^{(5)}$
are included.

In discussion, we would like to investigate the experiment realization
and validity of our theory. The nonlinearity of the material
is usually small and it is hard to observe the nonlinearity induced trapping
effect. But the ultra-high light intensity can induce enough permittivity change \cite{JiangNPhysics2010,UsmanSP2013,DeviPRA2017,GongPR2018optical,ZhangNL2018Nonlinearity-Induced,QuyOC2018,DeviEPL2019} and the potential well splitting using gold particle has been demonstrated with the aid of femtosecond pulse laser \cite{JiangNPhysics2010,ZhangNL2018Nonlinearity-Induced}. Thus we discussed metallic particles as examples in the study above,
where the particle permittivity is negative. It's noted that the high light intensity may damage the particle because of heating. However, we believe two factors will still promise the experimental verification. The first one is the material with no central-inversion symmetry. In this case, materials can have even order nonlinearity, such as $\chi^{(2)}$. As effects of $\chi^{(3)}$ has already been observed, material with $\chi^{(2)}$ and  $\chi^{(3)}$ can show the phenomena we discussed here within the reach. The second one is the developing material science and realization of meta-materials \cite{HuNPhotonics2019}. Artificial materials with desired high-order nonlinearity will demonstrate the theory here under lower light intensity in the future. Finally, we would like to mention that we discussed the materials with scalar nonlinearity coefficients in this work, which is true for homogeneous and isotropic materials. For anisotropic materials,
the nonlinearity coefficient is a tensor. The trapping will
then have more colorful phenomena deserved future investigations.

In summary, we investigated the features of the potential well of
an optically trapped particle with various orders of nonlinearity.
The potential well shape can have super-oscillation-like features
and Fano-resonance-like phenomenon, which provides finer particle
confinement far beyond the optical diffraction limit. It provides
a new method to manipulate micro/nano-particles parallel with sub-diffraction distance and may build a platform for high sensitivity measurement in opto-mechanical system.

Acknowledgment. We acknowledge the support from Ministry of Education,
Singapore (Grant No. R-263-000-D11-114) and discussions with Cheng-Wei Qiu. Y. Y. acknowledges support
of National Natural Science Foundation of China (11874102), Sichuan
Province Science and Technology Support Program (20CXRC0086). Y. Jiang acknowledges support from the Strategic Priority Research Program of Chinese Academy of Sciences (Grant no. XDA24020202) and  partial support from the National Natural Science Foundation of China (No. 11674389).

Disclosures. The authors declare no conflicts of interest.

\bibliography{NLpotential}

\bibliographyfullrefs{NLpotential}


\ifthenelse{\equal{\journalref}{aop}}{%
\section*{Author Biographies}
\begingroup
\setlength\intextsep{0pt}
\begin{minipage}[t][6.3cm][t]{1.0\textwidth} 
  \begin{wrapfigure}{L}{0.25\textwidth}
    \includegraphics[width=0.25\textwidth]{john_smith.eps}
  \end{wrapfigure}
  \noindent
  {\bfseries John Smith} received his BSc (Mathematics) in 2000 from The University of Maryland. His research interests include lasers and optics.
\end{minipage}
\begin{minipage}{1.0\textwidth}
  \begin{wrapfigure}{L}{0.25\textwidth}
    \includegraphics[width=0.25\textwidth]{alice_smith.eps}
  \end{wrapfigure}
  \noindent
  {\bfseries Alice Smith} also received her BSc (Mathematics) in 2000 from The University of Maryland. Her research interests also include lasers and optics.
\end{minipage}
\endgroup
}{}

\end{document}